\documentclass[aps,nofootinbib,showpacs,twocolumn,floatfix
]{revtex4}
\usepackage{graphicx}
\usepackage{epsf,amsfonts,amssymb,amsbsy}

\newcommand{\rt}{\rightarrow}

\newcommand{\Rsub}{\rm\scriptscriptstyle}

\begin{document}
\title{Extra charmonium states as bag-quarkonia}
\author{V.V.Kiselev}
\email{Valery.Kiselev@ihep.ru}
 \affiliation{Russian State Research
Center ``Institute for High
Energy Physics'', 
Pobeda 1, Protvino, Moscow Region, 142281, Russia\\ Fax:
+7-4967-742824\\
and\\ Moscow Institute of Physics and Technology, Institutskii
per. 9, Dolgoprudnyi, Moscow Region, 141701, Russia}
\begin{abstract}
Exotic states in the charmonium family are systematically treated
in the framework of simplest model with an effective coulomb-like
interaction of heavy quark and antiquark in the presence of static
excitation of quark-gluon modes responsible for a nonperturbative
term of potential, which provides with the confinement of quarks,
in terms of \textit{bag} over the threshold in the excitation
spectrum of vacuum fields. Once the spectrum has got quite a wide
mass gap, it allows us to approximate the bag contribution into
the potential by a constant value of bag mass at low distances
less than the bag size. The bag mass can be evaluated in a
constituent model. The analysis is given  for the bag contribution
into the distribution over the invariant mass of two pions in the
hadronic transition between the $S$-wave states of bag-quarkonium
and heavy quarkonium, that leads to the anomaly violating the
chiral limit in the region of low invariant masses, which agrees
with the observational data. Leptonic constants of vector states
are investigated in the presence of exotic states in the framework
of quasilocal sum rules. The extra states allow us to improve the
consistency of describing the measured widths of leptonic decays
for the complete set of vector states in the charmonium family.
\end{abstract}
\maketitle

\section{Introduction}
In the very beginning of charmonium era, it was quite spectacular
that the potential framework was not only qualitatively applicable
to the system of heavy quark and antiquark, but also
quantitatively successful in describing the mass spectrum of know
members of family as well as relevant static characteristics
determining mechanisms of producing the bound states and their
decays. The charmed quark is assigned to be heavy since its mass
is much greater than the scale characterizing the strong
interaction in the framework of Quantum Chromodynamics (QCD),
$m_\mathrm{Q}\gg\Lambda_\mathrm{QCD}$. The nonrelativistic motion
of heavy quarks with a velocity $v_\mathrm{Q}\ll 1$ provides with
the multipole expansion in QCD \cite{multipoleQCD,V2,Leut}, so
that in the leading order the static potential $V(r)$ can be
introduced. At short distances between the quarks, the potential
can be calculated in the perturbation theory
\cite{pert-V_stat-s,pert-V_stat-f,pert-V_stat-b,pert-V_stat-2},
while at large distances it can be approximated by the term giving
the quark confinement, i.e. the quark bounding inside colorless
states. We can mention the simplest example of such the potential,
the Cornell model \cite{cornell}. In this model the quarks in
color-singlet state are attracted due to the coulomb-like
interaction with an effective constant of ``single-gluon
exchange'' $\alpha_s^\mathrm{eff}$ and confined by the term
linearly raising with the distance increase. The slope of energy
increase $\sigma$ is consistent with the linear Regge trajectories
in the spectroscopy of hadrons composed of light quarks,
\begin{equation}\label{Cornell}
    V^\mathrm{Cornell}(r)=-\frac{4}{3}\,\frac{\alpha_s^\mathrm{eff}}{r}+\sigma\cdot
    r,
\end{equation}
where the factor of $\frac{4}{3}$ is caused by projecting to the
color-singlet state of quark-antiquark system. The value of
effective constant has been determined empirically by the
excitation spectrum in the charmonium family, while $\sigma\approx
0.18$ GeV$^2$. Average sizes $\langle r\rangle$ of charmonium
states are positioned in the intermediate region of transition
from the perturbative regime to the nonperturbative one, so that
the potential is quite accurately approximated by the logarithmic
function versus the distance \cite{log} with a dimensional
parameter $T$ determining an average kinetic energy in the heavy
quarkonium in accordance with the virial theorem\footnote{A weak
dependence of average kinetic energy on the heavy quark mass is
phenomenologically taken into account in the Martin model of
potential \cite{Martin}, functionally behaving as the power law
$r^k$ at $k\to 0$, that corresponds to a small perturbation of
logarithmic form in the region of approximation.}:
\begin{equation}\label{log}
    V(\langle r\rangle)\approx 2T\ln\frac{\langle
    r\rangle}{r_0},
\end{equation}
whereas $T\approx 0.38$ GeV, so that $T\sim\Lambda_\mathrm{QCD}$.

Therefore, a characteristic square of heavy quark momentum in such
the quarkonium is determined by the product of kinetic energy by
the reduced mass equal to $m_\mathrm{Q}/2$, i.e. $p^2\approx
m_\mathrm{Q} T$, hence, the square of heavy quark velocity is
equal to
\begin{equation}\label{velocity}
    v_\mathrm{Q}^2\sim \frac{p^2}{m_\mathrm{Q}^2}\sim \frac{T}{m_\mathrm{Q}},
\end{equation}
so it is suppressed in the case of heavy quarks. For example, the
charmed quark with the mass of $m_c\sim 1.6$ GeV has got
$$
    v_c^2\sim 0.25,
$$
and the potential model seems to be quite realistic, since
relativistic corrections (in the color-singlet state) have got a
magnitude of $v_\mathrm{Q}^2$.

The further progress in the development of potential approach was
related with the introduction of running coupling in QCD at small
distances in 1-, 2- and 3-loop order: the potentials by Richardson
\cite{Richardson}, Buchm\"uller and Tye \cite{Buch-Tye} and
Kiselev, Kovalsky and Onishchenko \cite{KKO,KKO-rus},
correspondingly, consistently with linearly raising term confining
the quarks. This way has allowed us to remove a discrepancy
between direct measuring the coupling of $\alpha_s$ and its value,
following from fitting the real spectrum of quarkonia in the
potential model with the running coupling constant only at the
3-loop approximation\footnote{The value of coupling constant is
inherently related with the splitting between the $1S$ and $2S$
levels in the quarkonia. At one- and two-loop orders the coupling
constant from the potential model has got a too high value beyond
the interval obtained by the direct measurement of $\alpha_s$.}.

However, the theoretical eligibility of potential description as
the framework seemed to become questionable, when one clarified
that the QCD vacuum has got a nontrivial structure, namely, it is
composed of quark-gluon condensates with the energy scale of the
order of $\Lambda_\mathrm{QCD}\sim 0.3$ GeV \cite{QCD-SR}.
Therefore, fluctuations of vacuum fields are characterized by the
time interval of $\tau_\mathrm{QCD}\sim 1/\Lambda_\mathrm{QCD}$,
while the relaxation time of quark-gluon fields in the potential
description is determined by the ratio of heavy quarkonium size to
the velocity of heavy quark motion,
\begin{equation}\label{tau-Q}
    \tau_\mathrm{Q}\sim\frac{r_\mathrm{Q\bar Q}}{v_\mathrm{Q}},
\end{equation}
so that by taking into account
$$
    r_\mathrm{Q\bar Q}\sim \frac{1}{m_\mathrm{Q}v_\mathrm{Q}},
$$
we get
\begin{equation}\label{tau2}
    \tau_\mathrm{Q}\sim \frac{1}{m_\mathrm{Q}v_\mathrm{Q}^2}.
\end{equation}
The vacuum fluctuations are irrelevant to the potential framework,
if the quarks can be considered as static and their interaction is
practically instantaneous in comparison with the influence of
nonperturbative effects, i.e.
\begin{equation}\label{tau3}
    \tau_\mathrm{QCD}\gg \tau_\mathrm{Q}\quad\Leftrightarrow\quad
    m_\mathrm{Q}v_\mathrm{Q}^2\gg \Lambda_\mathrm{QCD}.
\end{equation}
In other words\footnote{One could say that the potential would
change due to the vacuum fluctuations very slow in comparison with
a period of finite motion of heavy quarks, i.e. the potential is
static, indeed.}, the potential description can be considered as
rather reasonable and theoretically sound, if only the kinetic
energy of heavy nonrelativistic quarks is much greater the energy
scale fixing the quark-gluon condensates. This fact was recognized
by M.B.Voloshin\footnote{In fact, M.B.Voloshin actually formulated
the criterium of applicability for the multipole approximation in
QCD in the presence of vacuum fluctuations in the form of
condensates: $\tau_\mathrm{Q}/\tau_\mathrm{QCD}\ll 1$, while he
believed that the necessary constraint for the introduction of
static potential is the instantaneous interaction between the
nonrelativistic quarks, that oppositely requires
$\tau_\mathrm{Q}/\tau_\mathrm{QCD}\gg 1$, by his opinion, and
hence, it is evidently incompatible with the criterium of
multipole expansion, So, he suggested that the introduction of
static nonperturbative potential is principally impossible for
real charmed and beauty quarks except the exotic case of states in
vicinity of continuous threshold.} \cite{multipoleQCD,V2}, who
gave the estimate for a minimal acceptable value of heavy quark
mass, for which the description can be done in terms of potential
models with account of QCD condensates: $m_\mathrm{min} \sim 20$
GeV, which is in evident contradiction with the applicability of
potential models for the both charmed and bottom quarks, since
their kinetic energy has the same order of magnitude with the
scale $\Lambda_\mathrm{QCD}$, as we have seen above. Thus, the
motion of such heavy quarks could lead to the excitation of
nonzero vacuum fields from the condensates, so that the
interaction in the system evidently becomes nonstatic, which
contradicts with the success of potential framework in the
spectroscopy of charmonium and bottomonium.

The solution of this paradox comes from the quantum nature of
quark-gluon fields. The speculations presented above suggest that
the spectrum of excitations for the vacuum fields is continuous
and its threshold is determined by the energy scale of
$\Lambda_\mathrm{QCD}$, while the actual spectrum could by
essentially different. At first, the excitation spectrum for the
vacuum fields can begin with a wide mass gap, that is enough in
order to substantiate the applicability of potential framework to
the heavy quarkonium states below the threshold fixed by the mass
gap\footnote{The potential approach is completely substantiated in
the effective theory, the so-called potential nonrelativistic QCD
\mbox{(pNRQCD)} \cite{pNRQCD}.}. The threshold could be positioned
below the threshold for the pair production of hadrons with the
open charm (the production of D-meson pairs in the system of heavy
quark and antiquark), or it could be very close the continuous
threshold. Then, the second aspect becomes important: the spectrum
of excitations for the vacuum fields in the presence of heavy
quarkonium could be discrete in vicinity of mass gap, at least in
the sense of quasistationary description during the times
comparable with the period of heavy quark motion in the bound
state. Thus, it would be quite reasonable to introduce the notion
of ``bag'' for the lowest excited state of vacuum fields in the
system under consideration. The bag evidently has got no valence
degrees of freedom (isospin, charge etc. are equal to zero) and it
possesses the vacuum quantum numbers (spatial and charge parities
are positive) except the energy or mass. The existence of bag is
caused by the presence of valence quark and antiquark. In the
absence of valence quarks, the introduction of bag has no sense.
Therefore, the bag is not related with notion of glueballs or
hybrids as particles. In this respect, it would be incorrect to
associate the bag with  a separate quasistationary hadron-like
object, which interacts with other hadrons, for instance, with the
quarkonium, since the bag itself is not stationary without the
quarkonium. Thus, the system of bag-quarkonium differs from the
hadro-quarkonium recently introduced by M.B.Voloshin
\cite{Voloshin-rev,VY-hQQ}, since the system of hadro-quarkonium
is considered as the pair of stationary objects, which allow the
separated existence and interact a la the Van Der Waals forces
with each other.

\begin{figure}[th]
  \setlength{\unitlength}{1mm}
  \begin{center}
  \begin{picture}(75,51)
  \put(0,0){
  \includegraphics[width=70\unitlength]{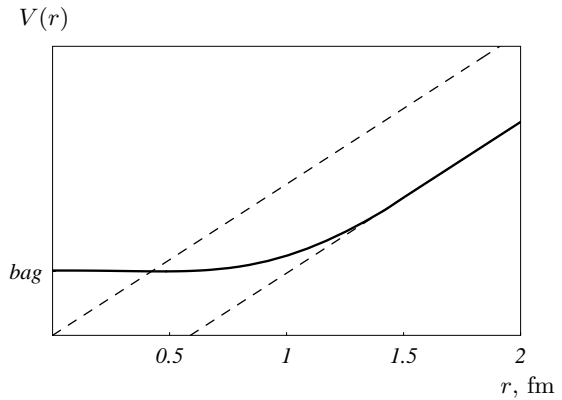}
  }
  \put(67,-3){$r$, fm}
  \put(3,46){$V(r)$}
  \end{picture}
  \end{center}
  \caption{The nonperturbative term in the potential energy due to
  the creation of bag in the system of two heavy quark and
  antiquark in the color-singlet state (the bright curve): the
  distance $r$ is measured in fm, the energy has arbitrary units,
  the static limit at zero distance between the quarks is marked
  by label \textit{bag}, the dashed lines correspond to linearly
  raising of potential at large distances in the presence of bag
  and in its absence. The bag size is ``chosen'' near 1 fm.}\label{fig-1}
\end{figure}

To my opinion, the influence of essentially nonperturbative
phenomenon called the bag on the system of heavy quarkonium is
reduced to the followings:
\begin{itemize}
    \item below the threshold of bag excitation, the heavy
    quarkonium permits the potential description, wherein the
    potential confining the quarks is phenomenologically
    approximated by the realistic term linearly raising with the
    distance increase;
    \item above the threshold of exciting the bag, the
    nonperturbative contribution is modified: if the heavy
    quarkonium size is less the size of bag $r_\mathrm{bag}$,
    then the bag presence is given by introducing the bag mass
    itself $E_\mathrm{bag}$ into the potential, while, if the
    quarkonium size becomes to exceed the bag size, then the
    linearly raising term of confining potential is activated
    again.
\end{itemize}
This situation is schematically shown in Fig. \ref{fig-1}, wherein
the nonperturbative potential of heavy quarks in the bag is
pictured. By the way, we assume that further excitations of vacuum
fields are separated from the bag by a mass gap, of course.

The shift of linearly raising asymptote of nonperturbatice term in
the presence of bag is determined by a interplay between its size
and mass. However, it is more essential that, if the size of heavy
quark-antiquark system is less than the bag size, i.e. when one
considers the lowest states of bag-quarkonium system, then the
interaction of heavy quark and antiquark ``to leading
approximation'' can be written in the form of following
\textit{potential}:
\begin{equation}\label{pot-eff}
    V^\mathrm{eff}(r)=-\frac{4}{3}\,\frac{\alpha_s^\mathrm{eff}}{r}+E_\mathrm{bag},
\end{equation}
where $\alpha_s^\mathrm{eff}$ is the effective constant of single
gluon exchange between the heavy quarks at the scale
characterizing the quark-antiquark system. Here we neglect the
linearly raising contribution in comparison with the bag mass.
Thus, we arrive to three-parametric spectral problem in the
nonrelativistic quantum mechanics: calculate the masses of states
in the system of bag-quarkonium with the coulomb-like attraction
at the given mass of heavy quark, effective constant
$\alpha_s^\mathrm{eff}$ and bag mass $E_\mathrm{bag}$.

In Section \ref{II} we present our treatment of extraordinary
states in the charmonium family in the framework of bag-quarkonium
system, that allows us to numerically estimate phenomenological
parameters from the empirical data. The parameters are positioned
in a region of quite reasonable values. In this way we refer to
the joint table of extra states of charmonia as given in
\cite{Godfrey-rev}: Tab. \ref{god}. We exclude the mesons with the
hidden strangeness $Y_s(2175)$ and beauty $Y_b$ as well as the too
broad state of $Y(4008)$ presented in the table, and remain them
beyond our consideration.

\begin{table*}[ht]
\begin{tabular}{|
l|c|c|c|c|c
|}

\colrule state    & mass $M$~(MeV) & width $\Gamma$~(MeV) &
$J^{PC}$ & decay modes      & production processes \\\hline
$Y_s(2175)$& $2175\pm8$&$ 58\pm26 $& $1^{--}$ & $\phi f_0(980)$ &
$e^+e^-$~(ISR),
$J/\psi$ decay \\
$X(3872)$& $3871.4\pm0.6$&$<2.3$& $1^{++}$ & $\pi^+\pi^-
J/\psi$,$\gamma J/\psi$
 & $B\rt KX(3872)$,
$p\bar{p}$ \\
$X(3875)$& $3875.5\pm 1.5$&$3.0^{+2.1}_{-1.7}$  &
&$D^0\bar{D^0}\pi^0$ & $B\rt
K X(3875)$ \\
$Z(3940)$& $3929\pm5$&$ 29\pm10 $& $2^{++}$ & $D\bar{D}$ &
$\gamma\gamma$        \\
$X(3940)$& $3942\pm9$&$ 37\pm17 $& $J^{P+}$ & $D\bar{D^*}$ &
$e^+e^-\rt J
/\psi X(3940)$ \\
$Y(3940)$& $3943\pm17$&$ 87\pm34 $&$J^{P+}$ & $\omega J/\psi$ &
$B\rt K Y(39
40)$          \\
$Y(4008)$& $4008^{+82}_{-49}$&$ 226^{+97}_{-80}$ &$1^{--}$&
$\pi^+\pi^- J/\psi$
& $e^+e^-$(ISR)  \\
$X(4160)$& $4156\pm29$&$ 139^{+113}_{-65}$ &$J^{P+}$&
$D^*\bar{D^*}$& $e^+e^-\rt
 J/\psi X(4160)$\\
$Y(4260)$& $4264\pm12$&$ 83\pm22$ &$1^{--}$&  $\pi^+\pi^- J/\psi$
& $e^+e^-$(ISR
)       \\
$Y(4350)$& $4361\pm13$&$ 74\pm18$ &$1^{--}$&  $\pi^+\pi^-
\psi^{\prime}$ & $e^+e
^-$(ISR)       \\
$Z(4430)$& $4433\pm5$&$ 45^{+35}_{-18}$ & ? &
$\pi^{\pm}\psi^{\prime}$ & $B\rt K
Z^{\pm}(4430)$\\
$Y(4660)$& $4664\pm12$&$ 48\pm15 $ &$1^{--}$&  $\pi^+\pi^-
\psi^{\prime}$
& $e^+e^-$(ISR)       \\
$Y_b$     & $\sim 10,870$ & ?    &  $1^{--}$ &
$\pi^+\pi^-\Upsilon(nS)$
& $e^+e^-$       \\
\hline
\end{tabular}
\caption{The joint table of cadidates in members of charmonium
family: mesons $XYZ$, as copied from \cite{Godfrey-rev}. }
\label{god}
\end{table*}

The modification of hadronic transitions from the bag-quarkonium
states to the ordinary quarkonium with emission of pion pair is
investigated in Section \ref{III}. The presence of bag involves a
new term in the amplitude in comparison with transitions between
the ordinary S-wave states of heavy quarkonium, that changes the
distribution versus the invariant mass of two pions, i.e. it can
serve as the reason for the observed anomalous behavior of
distribution in transitions of $\psi(3770)\to J/\psi\pi\pi$ and
$\Upsilon(3S)\to \Upsilon(1S)\pi\pi$. This fact is the argument in
favor of our assignment of $\psi(3770)$ and $\Upsilon(3S)$ as
basic vector states in the system of bag-charmonium and
bag-bottomonium, respectively.

Section \ref{IV} is devoted to the analysis of leptonic constants
of vector states in the charmonium family in the framework of
quasilocal sum rules \cite{quasilocal-SR}. The study shows that
widths of observed leptonic decays of vector states do not
contradict with introducting the extra states of bag-charmonium.
Moreover, the consistency of predictions with the experimental
data is systematically improved.

In Conclusion we summarize the obtained results and discuss their
possible implications to the study of mechanisms for the
production and decays of extra heavy-quarkonium states.

\section{Analysis of exotica in the spectrum of charmonium \label{II}}

The spectrum of energy in the problem with coulomb interaction is
well known. For the system with the reduced mass
$m_\mathrm{red}=\frac{1}{2}m_\mathrm{Q}$, the spin-dependent term
due to effective single-gluon exchange takes the form
\begin{widetext}
\begin{equation}\label{SD}
\begin{array}{ccl}
    V_\mathrm{SD}&\hskip-5pt=&\hskip-3pt\displaystyle
    \frac{4}{3}\,\alpha_s^\mathrm{eff}\,\frac{8\pi}{3}\,
    \frac{1}{m_\mathrm{Q}^2}\,(\boldsymbol s_1\cdot\boldsymbol
    s_2)\,\delta(\boldsymbol r)
    +
    2\alpha_s^\mathrm{eff}\,\frac{1}{m_\mathrm{Q}^2}\,(\boldsymbol
    L\cdot\boldsymbol S)\,\frac{1}{r^3}
+
    \frac{4}{3}\,\alpha_s^\mathrm{eff}\frac{1}{m_\mathrm{Q}^2}\,
    \left\{3(\boldsymbol s_1\cdot\boldsymbol n)(\boldsymbol s_2\cdot
    \boldsymbol n)-(\boldsymbol s_1\cdot\boldsymbol s_2)\right\}
    \,\frac{1}{r^3},
\end{array}
\end{equation}
\end{widetext}
In (\ref{SD}) the first term corresponds to spin-spin ``contact''
interaction of quark and antiquark, the second represents the
spin-orbital interaction, the third gives the tensor forces. Here
$\boldsymbol r=\boldsymbol r_1-\boldsymbol r_2$ is the relative
distance between the quarks, the unit vector is $\boldsymbol
n=\boldsymbol r/r$, $\boldsymbol L$ denotes the summed orbital
momentum of quarks, and  $\boldsymbol S=\boldsymbol
s_1+\boldsymbol s_2$ gives the summed spin. The masses of levels
$n^{2S+1}L_J$ with $n$ and $J$ being the principal quantum number
and total momentum are given by
\begin{widetext}
\begin{equation}\label{M}
\begin{array}{ccl}
    M_\mathrm{bag-Q\bar Q}[n^{2S+1}L_J]&=&\displaystyle
    2m_\mathrm{Q}+E_\mathrm{bag}-\frac{m_\mathrm{Q}}{4n^2}\,
    \left(\frac{4}{3}\,\alpha_s^\mathrm{eff}\right)^2
+
    \frac{m_\mathrm{Q}}{6n^3}\,\left(\frac{4}{3}\,\alpha_s^\mathrm{eff}\right)^4
    \left\{S(S+1)-\frac{3}{2}\right\}\,\delta_{L0}\\[4mm]
    &+&\displaystyle
    \frac{m_\mathrm{Q}}{n^3}\,\left(\frac{4}{3}\,\alpha_s^\mathrm{eff}\right)^4
    \,\frac{1}{(2L+1)\big((2L+1)^2-1\big)}\,
    \raisebox{1pt}{$\scriptstyle\times$}
    \\[5mm]
    &&\displaystyle
    \left\{\frac{3}{2}\,(\boldsymbol L\cdot\boldsymbol S)-\frac{1}
    {4L(L+1)-3}\,\left(3(\boldsymbol L\cdot\boldsymbol
    S)^2+\frac{3}{2}\,(\boldsymbol L\cdot\boldsymbol S)-\boldsymbol L^2\boldsymbol
    S^2\right)
    \right\},
\end{array}
\end{equation}
\end{widetext}
whereas
$$
\begin{array}{rcl}
    (\boldsymbol L\cdot\boldsymbol S)&\hskip-3pt=&\hskip-3pt\frac{1}{2} \big\{
    J(J+1)-L(L+1)-S(S+1)\big\},\\[2mm]
    \boldsymbol L^2&\hskip-3pt=&\hskip-3pt L(L+1),\quad
    \boldsymbol S^2=S(S+1).
\end{array}
$$
The spatial and charge parities, respectively given by
$P=(-1)^{L+1}$ and $C=(-1)^{S+L}$. Particularly, at $L=0$, the
masses of vector states $J^{PC}=1^{--}$ are equal to
\begin{equation}\label{M1--}
    \begin{array}{ccl}
    M_\mathrm{bag-Q\bar Q}[n^3S_1]&\hskip-3pt=&\hskip-3pt\displaystyle
    2m_\mathrm{Q}+E_\mathrm{bag}-\frac{m_\mathrm{Q}}{4n^2}\,
    \left(\frac{4}{3}\,\alpha_s^\mathrm{eff}\right)^2\\[4mm]
    &\hskip-3pt+&\hskip-3pt\displaystyle
    \frac{m_\mathrm{Q}}{12n^3}\,\left(\frac{4}{3}\,\alpha_s^\mathrm{eff}\right)^4,
\end{array}
\end{equation}
while at $L=1$, the masses of spin-triplet $J^{++}$-states are
equal to
\begin{equation}\label{MJ++}
\begin{array}{ccl}
    \hskip-3ptM_\mathrm{bag-Q\bar Q}[n^3P_J]&\hskip-4pt=&\hskip-2pt\displaystyle
    2m_\mathrm{Q}+E_\mathrm{bag}-\frac{m_\mathrm{Q}}{4n^2}\,
    \left(\frac{4}{3}\,\alpha_s^\mathrm{eff}\right)^2\\[6mm]
    &\hskip-17pt+&\hskip-12pt\displaystyle
    \frac{m_\mathrm{Q}}{24n^3}
    \,\left(\frac{4}{3}\,\alpha_s^\mathrm{eff}\right)^4\,
    \left\{
    \begin{array}{cl}
    +\frac{7}{5}, & J=2,\\[2mm]
    -1, &J=1,\\[2mm]
    -4, &J=0.
    \end{array}
    \right.
\end{array}
\end{equation}
The review of charmonium family is presented in
\cite{Voloshin-rev,Godfrey-rev}. We assign $\psi(3770)$ as the
basic extraordinary vector state, which is distinguished by the
anomalous distribution in the spectrum of invariant masses for two
pions in the transition $\psi(3770)\to J/\psi \pi\pi$. One usually
considers that this state  is positioned in vicinity of $1D$ level
of quark-antiquark system $c\bar c$, and its nonzero width of
decay into lepton pair is caused by  a mixing with the nearest
$S$-wave level $\psi^\prime=\psi(2S)=\psi(3680)$. Then, one can
treat it as the ordinary state, and the anomaly mentioned has a
reason, which is, for instance, an occasional suppression of
dominant term in the creation of pion pair (see the analysis in
\cite{Voloshin-rev}). However, in Section \ref{III} we argue in
favor of that the anomalous distribution versus the invariant mass
of two pions is the natural consequence, if one assigns
$\psi(3770)$ as the system of bag-charmonium. Therefore, we
identify $\psi(3770)$ with the state $1^{--}$ in the system of
bag-charmonium with the quantum numbers $1^3S_1$.

\begin{figure}[th]
\setlength{\unitlength}{1mm}
  \begin{center}
  \begin{picture}(77,51)
  \put(5,0){
  \includegraphics[width=70\unitlength]{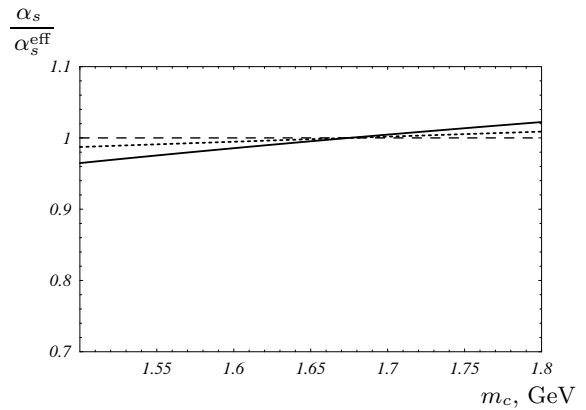}
  }
  \put(65,-3){$m_c$, GeV}
  \put(2,46){$\displaystyle\frac{\alpha_s\,}{\,\alpha_s^\mathrm{eff}}$}
  \end{picture}
  \end{center}
  \caption{The ratio of running coupling constant in QCD to
  the effective constant, calculated in accordance with
  (\ref{DeltaM}) and (\ref{effect}): the 1-loop approximation is given by
  the solid line, the 2-loop curve is show by dotted line, the dashed line
  refers to the value equal to  1.}\label{fig-2}
\end{figure}

The other benchmark of exotic state is $X(3872)$, which has got
the definite quantum numbers $1^{++}$. This state has to be
identified with the first excitation at the same quantum numbers
in the system of bag-charmonium, i.e. $2^3P_1$. Thus, in order to
determine the parameters of system we have got two levels with the
mass difference
\begin{equation}\label{DeltaM}
    \Delta
    M=\frac{3m_c}{16}\,\left(\frac{4}{3}\,\alpha_s^\mathrm{eff}\right)^2
    \left\{1-\frac{17}{36}\,\left(\frac{4}{3}\,\alpha_s^\mathrm{eff}\right)^2\right\},
\end{equation}
that numerically gives $\Delta M=102$ MeV. Eq. (\ref{DeltaM})
allows us to write down the effective constant as the function of
charmed quark mass
\begin{equation}\label{effect}
    \left(\frac{4}{3}\,\alpha_s^\mathrm{eff}\right)^2=\frac{18}{17}
    \left\{1-\sqrt{1-\frac{1802}{27}\,\frac{\Delta
    M}{m_c}}\right\}.
\end{equation}
Note that the obtained behavior of effective constant versus the
charmed quark mass in the limits from 1.3 to 1.8 GeV is in a good
agreement with the 1-loop dependence of running coupling constant
in QCD
\begin{equation}\label{run1}
    \alpha_s(\mu)=\frac{2\pi}{\beta_0\ln[\mu/\Lambda_{QCD}]},
\end{equation}
if one puts that the scale $\mu$ is determined by the
characteristic momentum of heavy quark in the heavy quarkonium,
i.e.
\begin{equation}\label{run-mu}
    \mu^2=m_\mathrm{Q}\,T,
\end{equation}
wherein $T\approx 0.4$ GeV,  $\Lambda_\mathrm{QCD}\approx 190$
MeV, and
$$
    \beta_0=11-\frac{2}{3}\, n_\mathrm{f}
$$
at the number of quark flavors contributing to the renormalization
for the scales in problem, $n_\mathrm{f}=3$. The fact of
consistency between the empirical dependence of (\ref{effect}) and
renormalization group (RG) one of (\ref{run1})--(\ref{run-mu}) is
illustrated in Fig. \ref{fig-2}, wherein the ratio of RG constant
to the effective one is shown after the normalization at
$m_c\approx 1.65$ GeV.

\noindent The figure shows that in the wide interval of pole mass
for the charmed quark, the effective constant coincides with the
RG running constant within the accuracy better than 5\%. The
agreement between these two values becomes more spectacular in the
2-loop approximation for the running coupling constant in QCD
$$
    \alpha_s^\mathrm{[2-\ell]}(\mu)=\frac{2\pi}{\beta_0\ln[\mu/\Lambda_{QCD}]}
    \left(1-\frac{\beta_1}{\beta_0^2}\,\frac{\ln\ln[\mu^2/\Lambda_{QCD}^2]}{
    \ln[\mu/\Lambda_{QCD}]}
    \right),
$$
where $\beta_1=51-19n_\mathrm{f}/3$, after the appropriate
correction of $\Lambda_\mathrm{QCD}$. However, the usage of 2-loop
approximation is evidently beyond the accuracy of consideration,
since the model potential does not take into account the
dependence of effective constant versus the distance in the second
order over $\alpha_s$.

Nevertheless, we can establish that the running coupling constant
of QCD at the scale fixed by the heavy quark mass and its kinetic
energy slowly depending on the mass, allows us to get the
empirical value for splitting the masses of lightest vector and
pseudovector states in the system of bag-charmonium.

\subsection{The pole mass of heavy quark}

\begin{figure}[th]
\setlength{\unitlength}{1mm}
  \begin{center}
  \begin{picture}(77,47)
  \put(5,0){
  \includegraphics[width=70\unitlength]{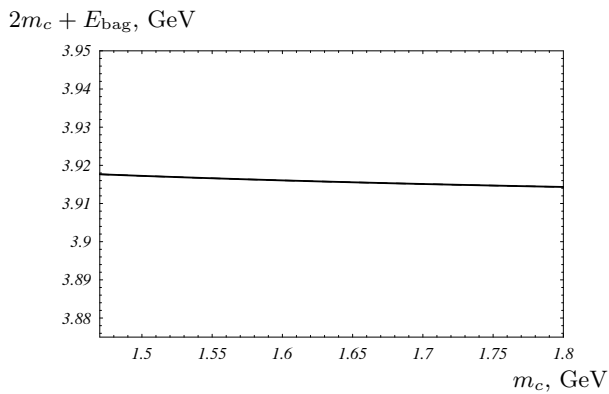}
  }
  \put(67,-3){$m_c$, GeV}
  \put(0,45){$2m_c+E_\mathrm{bag}$, GeV}
  \end{picture}
  \end{center}
  \caption{The sum of bag mass and double mass of charmed quark
  versus the pole mass as the result of fitting the masses of
  lightest vector and pseudovector states of bag-charmonium in
  accordance with (\ref{M1--})--(\ref{effect}).}\label{fig-3}
\end{figure}

It is spectacular (see Fig. \ref{fig-3}) that the sum of pole
masses for charmed quark and antiquark with the bag mass is
practically independent of the pole mass of charmed quark with the
accuracy better than 5 MeV,
\begin{equation}\label{const}
    \mathcal{E}_\mathrm{IRstab}=2m_c+E_\mathrm{bag}=\mbox{const.},
\end{equation}
after the implication of (\ref{M1--})--(\ref{effect}).

Numerically, we find
\begin{equation}\label{const2}
    \mathcal{E}_\mathrm{IRstab}\approx 3915\mbox{ MeV.}
\end{equation}
The constant value of this quantity points to that the infrared
instability of heavy-quark pole mass is exactly cancelled by the
infrared instability in the static potential approximated by the
dominant contribution of bag mass in the case under consideration.
The instability of pole mass appears in the perturbation theory of
QCD in the form of renormalon
\cite{Dyson,tHooft,Parisi,Lautrup,Mueller,Zakharov,Beneke}, that
exhibits the factorial growth of coefficients in the perturbative
series as dictated by the presence of infrared pole in the
coupling constant of QCD at the scale of $\Lambda_\mathrm{QCD}$.
Thus, the observed quantity being the mass of bound state for the
two heavy quarks and bag is actually stable in the infrared
region, that we mark by the subscript in (\ref{const}). The value
of charmed quark mass itself is completely determined by the
separation of bag mass in the approximation considered.

In this respect it would be interesting to make the following
analogy: at first, consider the light quarkonium with the ordinary
$u$- and $d$-quarks instead of the heavy quarkonium. Let us
neglect the light quark masses as well as the energy of their
kinetic motion, if there is no orbital rotation. In the basic
$S$-wave state the mass of such the quarkonium with no account of
spin-dependent interactions is determined by the bag mass
\begin{equation}\label{1S}
    m(1S)\approx E_\mathrm{bag},
\end{equation}
i.e. the state itself is the bag with the valence quantum numbers
given by the light quarks. In the constituent framework the bag
energy fixes the constituent mass of light quark by
$m_\mathrm{const.}\approx\frac{1}{2}E_\mathrm{bag}$. Such the
bag-quarkonium after the introduction of spin-dependent forces is
observed as the pseudoscalar $\pi$- and vector $\rho$-mesons.
Therefore, the standard procedure of cancelling the contribution
of spin-dependent energy leads to
$$
    m(1S)\approx\frac{1}{4}(3m_\rho+m_\pi)\approx 612\mbox{ MeV,}
$$
that gives the approximate value of bag mass in accordance with
(\ref{1S}).

Second, the same way can be applied to the consideration of
``constituent model'' of quarkonium composed of light and heavy
quarks. Then, in the basic $S$-wave state with no account for the
spin-dependent forces, such the heavy-light bag-quarkonium has got
the mass equal to
\begin{equation}\label{1S-Qq}
    M_\mathrm{q\bar Q}[1S]\approx
    m_\mathrm{Q}+m_\mathrm{const.}\approx
    m_\mathrm{Q}+\frac{1}{2}\,E_\mathrm{bag}.
\end{equation}
Particularly, for the basic level of D-mesons we get
\begin{equation}\label{D}
    M_\mathrm{D}[1S]\approx\frac{1}{2}\,\mathcal{E}_\mathrm{IRstab}\approx
    1958\mbox{ MeV,}
\end{equation}
that should be compared with the experimental value averaged over
the spins for neutral D-mesons, for instance,
\begin{equation}\label{D-exp}
    M_\mathrm{D}^\mathrm{exp}[1S]=\frac{1}{4}\, (3 m_{D^*}+m_D)\approx
    1972\mbox{ MeV.}
\end{equation}
Comparing (\ref{D-exp}) with (\ref{D}) shows that the speculations
in the framework of constituent model or bag-quarkonium allow us
to make the calculations with the accuracy about 15 MeV,
characteristic for the potential models in general.

\begin{figure}[th]
\setlength{\unitlength}{1mm}
  \begin{center}
  \begin{picture}(77,47)
  \put(5,0){
  \includegraphics[width=70\unitlength]{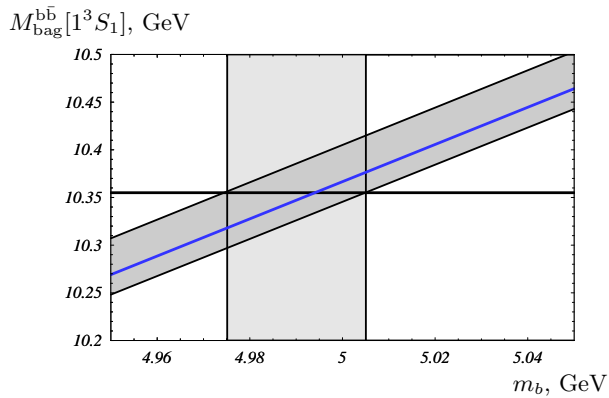}
  }
  \put(67,-3){$m_b$, GeV}
  \put(0,45){$M_\mathrm{bag}^\mathrm{b\bar b}[1^3S_1]$, GeV}
  \end{picture}
  \end{center}
  \caption{The calculation of mass for the basic vector state in
  the system of bag-bottomonium. The vertical band shows the
  permitted values of $b$-quark pole mass, the horizontal line
  marks the experimental value of $\Upsilon(3S)$ mass, the sloped band
  restricts the prediction with the variation of effective
  constant in limits $0.423<\frac{4}{3}\,\alpha_s^\mathrm{eff}<0.484$
  in the bag-bottomonium system, the bright sloped line gives the
  prediction for the mass at the constant obtained by the
  rescaling of its value from the value for the bag-charmonium in
  accordance with the renormalization group law.}\label{fig-4}
\end{figure}

Thus, in the framework under study we prefer for prescribing the
pole mass of charmed quark to the value equal to
\begin{equation}\label{m_c}
    m_c\approx
    \frac{1}{2}\,\big[\mathcal{E}_\mathrm{IRstab}-m(1S)\big]\approx
    1650\pm 15\mbox{ MeV,}
\end{equation}
and analogously, fixing the pole mass of bottom quark by the value
equal to
\begin{equation}\label{m_b}
\begin{array}{rcl}
    m_b&\approx& m_\mathrm{B}^\mathrm{exp}[1S]-\frac{1}{2}\,m(1S)\\[2mm]
    &\approx&
    m_c+m_\mathrm{B}^\mathrm{exp}[1S]-m_\mathrm{D}^\mathrm{exp}[1S]\\[2mm]
    &\approx & 4995\pm 15\mbox{ MeV,}
\end{array}
\end{equation}
which surprisingly agree with the values obtained in the framework
of analysis performed for charmonium and bottomonium described by
the dominant coulomb interaction with account for corrections both
in the perturbation theory and due to the quark-gluon condensates
\cite{TY1,TY2,Y3,Y+}. Then, after the rescaling the effective
constant of coulomb-like interaction by the renormalization group
law to the system of bottomonium, we can calculate the mass of
basic vector state of bag-bottomonium\footnote{Once such the basic
vector state coincides with the position of $\Upsilon(3S)$,
possessing the anomalous properties in two-pion transition into
the low-lying vector levels of bottomonium, while its first
excitation is positioned at $\Upsilon(4S)$ with the prediction
accuracy of 30 MeV, though the splitting between the further
excitations becomes less than the uncertainty of prediction, so
that taking into account the width of $\Upsilon(4S)$ comparable
with the uncertainty of prediction, we do not make certain
statement concerning for the definite correspondence of any exotic
state to the accepted notation of $\Upsilon(4S)$.}, as shown in
Fig. \ref{fig-4}:
\begin{equation}\label{bottom}
    M_\mathrm{bag}^\mathrm{b\bar b}[1^3S_1]\approx 10355\pm 30\mbox{ MeV.}
\end{equation}

Note, that the same result is achieved, if we vary the pole mass
of charmed quark in the limits from 1.5 to 1.7 GeV with the
consistent change of $b$-quark pole mass as well as the value of
effective constant for the coulomb-like interaction.

At the pole masses fixed above, the coupling constants are equal
to
\begin{equation}\label{coupl}
    \alpha_s^\mathrm{eff}[\mathrm{c\bar c}]\approx 0.479,\qquad
    \alpha_s^\mathrm{eff}[\mathrm{b\bar b}]\approx 0.347.
\end{equation}
Characteristic sizes of heavy quark and antiquark system are
determined by the ``Bohr radius''
\begin{equation}\label{bohr}
    a=\frac{3}{2m_\mathrm{Q}\alpha_s^\mathrm{eff}},
\end{equation}
which is, for example, approximately equal to $a\approx 0.38$ fm
at $m_c=1.65$ GeV. Remember, that, as well known, the wave
functions of states bound in the coulomb potential exponentially
decline versus the distance with the damping lengths of $a\,n$,
where $n$ is the principal quantum number, so that for the
bag-charmonium this length is less than 1 fm at $n=1,2$. Supposing
that the bag itself has the size somewhat greater than 1 fm, we
draw the conclusion that the approximation introduced for the
potential as the sum of coulomb term and bag mass is quite
applicable for the low-lying levels of bag-charmonium.

\subsection{States at $n=1,2$}

Fixing the model parameters by means of considering the splitting
between the two lowest vector and pseudovector levels of
bag-charmonium at $n=1,2$ allows us to make spectroscopic
predictions for further members of family with the same values of
principal quantum number. The uncertainty of such the prediction
is about 5 MeV. The state masses are listen in Tab. \ref{tab-1}.
Emphasize that for the coulomb-like states the splitting of levels
due to the forces depending on the quark spins approximately
decreases with the growth of principal quantum number by the
scaling law $1/n^3$, that reasonably agrees with the data in Tab.
\ref{tab-1}.
\begin{table}[th]
  \centering
  \begin{tabular}{|c|c|c|}
    \hline
    $J^{PC}$ & $n=1$ & $n=2$ \\
    \hline
    $0^{-+}$ & 3678 & 3865 \\
    $1^{--}$ & 3770 & 3876 \\
    $2^{++}$ & - & 3875 \\
    $1^{++}$ & - & 3872 \\
    $0^{++}$ & - & 3868 \\
    $1^{+-}$ & - & 3873 \\
    \hline
  \end{tabular}
  \caption{Masses of bag-charmonium at $n=1,2$ in MeV.}\label{tab-1}
\end{table}

Two of listened states have been associated with the following
ones:
$$
    1^{--}[1^3S_1]\mapsto \psi(3770),\qquad
    1^{++}[2^3P_1]\mapsto X(3872),
$$
the quantum numbers $J^{PC}$ of which are established
experimentally, while the exotic level $X(3875)$ with unidentified
quantum numbers could get the two following prescriptions:
$$
    X(3875)\,\raisebox{-2mm}{$\stackrel{\mbox{\large
    $\nearrow$}}{\mbox{\large$\searrow$}}$}\;
    \begin{array}{cc}
    1^{--}[2^3S_1],\\[4mm]
    2^{++}[2^3P_2].
    \end{array}
$$

\subsection{States at $n>2$}
The accepted approximation for the potential as the sum of coulomb
attraction and bag mass is valid at $n=1,2$, but it can be
modified in the case of higher excitations because of perturbation
responsible for a ``soft'' transition into the regime of
confinement, since at $n=3$ the system composed of charmed quark
and antiquark has got the size about 1 fm, i.e. it is close to the
confinement scale, and hence, to the bag size itself. Therefore,
we have to account for some other nonperturbative terms.

So, the presence of gluon condensate, i.e. an external field,
leads to the contribution caused by the second order of
perturbation theory due to the chromoelectric dipole interaction
to the next-to-leading order in the velocity of heavy quarks, that
has the form\footnote{We deal with the general formula for the
correction to the energy $\delta M\sim \langle V\rangle^2/\delta
E$ with the perturbation due to the electric dipole $V\sim g_s r
\mathcal E$, so that the charge squared gives the coupling
constant $g_s^2\mapsto \alpha_s$, the square of chromoelectric
field is reduced to the square of gluon stress
$\mathcal{E}^2\mapsto - G_{\mu\nu}^2$, and the characteristic size
gives the Bohr radius of $n$-th excitation $r\mapsto a\,n$.
Actually, the strict calculation of nonperturbative term due to
the gluon condensate as well as the complete analysis of
nonrelativistic quark-antiquark system with coulomb interaction
was done in \cite{TY1,TY2,Y3,Y+} for the charmonium and bottmonium
in the framework of multipole expansion of QCD
\cite{multipoleQCD}. The authors used exact formulae for the Green
function of color-octet state, so that $\delta
M_{nl}=n^6m_\mathrm{Q}\langle\alpha_s G^2\rangle\,a^4 \pi
\epsilon_{nl}/16$. At first, this correction gives sixth power of
principal quantum number, in fact, because of averaging $\langle
r\rangle\mapsto a n^2$ valid for the coulomb potential (the exact
formula $\langle r\rangle=a[3n^2-l(l+1)]/2$ includes the
dependence on the orbital momentum). Second, the numerical
dimensional factor in front of correction is certainly definite in
such the estimate. Third, one gets the opportunity to take into
account for the orbital quantum number in terms of factor
$\epsilon_{nl}\approx 1.5$, which value is also strictly known.
However, we do not transfer such the analysis to the case of
considering the bag-quarkonium, because we suggest that the bag
essentially distorts the propagation of color-octet state, since
it cuts off the wave functions in infrared, especially in the case
of subtraction between the quark and antiquark. In this respect,
our approach to the problem is less strict, but, to our opinion,
it is more close to the actual physical situation, so that the
fourth power in the dependence of correction on the principal
quantum number is more realistic, while the numerical dimensional
factor is fitted phenomenologically, because the modification of
wave functions or Green function in infrared can essentially
``renormalize'' this factor.}
\begin{equation}\label{pert}
    \delta M_n 
    =\kappa_n (a\,n)^2
\end{equation}
at
$$
    \kappa_n\sim A\,\frac{\langle \alpha_s
    G^2\rangle}{\delta E_n},
$$
where $\langle \alpha_s G^2\rangle\sim\Lambda_\mathrm{QCD}^4$ is
the gluon condensate, $\delta E_n\sim
m_\mathrm{Q}(\frac{4}{3}\alpha_s^\mathrm{eff})^2/4n^2$ is the
characteristic energy of binding the heavy quarks, and $A$ denotes
the numerical dimensionless factor caused by color and spatial
effects. Then, at $\Lambda_\mathrm{QCD}\sim 0.2$ GeV in the case
of bag-charmonium we find the estimate $\kappa_n\sim A\,n^2
m_\mathrm{Q}a^2\langle\alpha_s G^2\rangle\sim
A\,n^2\raisebox{1pt}{$\scriptstyle\times$}10^{-3}\mbox{ GeV}^3$.
Remember that the positive value of gluon condensate corresponds
to ``negative'' square of chromoelectric field; the intermediate
state of quark-antiquark pair is the color octet with the
subtraction between the quarks, hence, it has the energy greater
than the bound states under study, that leads to positive value of
correction to the mass of state\footnote{Correction (\ref{pert})
can be roughly described by introducing the perturbation potential
$V=\kappa^\prime r^2$, so that for the coulomb functions of
initial states the shift of energy is equal to
\begin{equation}\label{pert2}
    \langle V\rangle=\kappa^\prime\,\frac{1}{2}\,a^2n^2[5n^2+1-3L(L+1)],
\end{equation}
which repeats the general behavior versus the principal quantum
number, of course. Eq. (\ref{pert2}) supposes that the correction
to the energy is determined by one and the same value of parameter
$\kappa^\prime$ independent of the state, that is certainly too
strong suggestion. Though, the final expression can be treated as
the factorization of overall factor, while the residual dependence
versus the parameters of state is close to the fourth power of
principal quantum number. In (\ref{pert2}) one could account for
correction variation because of nonzero value of orbital momentum,
but this piece of subtlety is evidently beyond the accuracy
suggested in the derivation of correction itself.}. In addition,
the effective constant related with the gluon condensate can
depend on the fact that the chromoelectric field is partially
screened by the bag, which is the infrared object itself. This
fact can lead to a dependence of effective value related with the
condensate, that has the step-like form changing the value at a
critical distance given by the bag size, so that
$
    \kappa\mapsto\kappa\big(1+\varepsilon\vartheta[n_\star-n]\big).
$

Let us evaluate the effect of perturbation introduced in the case
of neglecting the spin-dependent forces\footnote{As we have
already mentioned above, the splitting of higher excitations is
significantly suppressed by the factor of $1/n^3$, so that the
variation of estimates due to the spin-dependent forces at $n=3$
is less than 3 MeV, that is certainly below the uncertainty in the
formula for the interpolation used.} by setting
\begin{equation}\label{kappa1}
    \kappa_n=n^2\kappa,\qquad
    \begin{array}{ccl}
    \kappa&=&\kappa_0\big(1+\varepsilon\vartheta[n_\star-n]\big),\\[1mm]
    \kappa_0&=&1.1\raisebox{1pt}{$\scriptstyle\times$}10^{-4}\mbox{
    GeV}^3,\\[1mm]
    \epsilon&=&0.3,\quad
    4<n_\star<5.
    \end{array}
\end{equation}
{The value of $\kappa_0$ is in fact fixed by the position
of higher excitations. It is important that the functional
dependence of correction on the principal quantum number truly
reflects the structure of extraordinary states in the charmonium
family. The introduction of nonzero value for the parameter
$\varepsilon$ allows us to slightly displace the positions of
levels at $n=3,4$ by 10 and 30 MeV, respectively, that is
comparable with the accuracy calculation of methods generally
involved in the framework of potential models. Nevertheless, the
physical meaning reflected by parameters $\varepsilon$ and
$n_\star\approx 4.5$ is quite clear: the degree of screening the
chromoelectric field by the bag is of the order of 30 \%, and the
bag size is about $a\,n_\star\approx 1.5-1.6$ fm.}

Then, we find that at $n=1,2$ the contribution of perturbation is
less than 1 MeV, while for the higher excitations we get
predictions corresponding to exotic states of charmonium:
\begin{equation}\label{predict}
\begin{array}{ccl}
    M_\mathrm{bag}^\mathrm{c\bar c[n=3]}
    =3939
    \mbox{ MeV}&\mapsto & \begin{array}{c}
    X(3940),\,Y(3940),\\
    Z(3940),\end{array}\\[4mm]
    M_\mathrm{bag}^\mathrm{c\bar c[n=4]}=4037
    \mbox{ MeV}&\mapsto & \psi(4040),\\[2mm]
    M_\mathrm{bag}^\mathrm{c\bar c[n=5]}=4156
    \mbox{ MeV}&\mapsto & \psi(4160),\,X(4160),\\[2mm]
    M_\mathrm{bag}^\mathrm{c\bar c[n=6]}=4424
    \mbox{ MeV}&\mapsto & \psi(4415).
\end{array}
\end{equation}
Hypothetic higher excitations can suffer from the influence of
linearly raising term in the potential, so that their masses could
exceed the naive estimate at $n>5$, say, for
$M_\mathrm{bag}^\mathrm{c\bar c[n=6]}\approx 4.42$ GeV (the size
of quar-antiquark system exceeds 2 fm), but such the high values
of excitation energy signal that thresholds of some other effects
could open. Particularly, the excitation of bag could take place,
that we discuss in the next subsection.

\subsection{Exciting the bag}

The method of estimating the mass of basic state for the bag in
the framework of constituent model for $\pi$- and $\rho$-mesons
suggests that the bag could have excitations. The first excitation
is related with the system of $a$-mesons. The spin-average state
has the mass\footnote{We use the prescription of $n_rl$, where
$n_r$ is the radial quantum number.} equal to
\begin{equation}\label{excite1}
    m(1P)=\frac{1}{9}\big(5 m_{a_2}+3m_{a_1}+m_{a_0})\approx
    1260\mbox{ MeV.}
\end{equation}
However, in contrast to the $S$-wave state, the presence of
$P$-wave suggests that the constituent quarks get the kinetic
energy of orbital rotation,
\begin{equation}\label{orbit}
    E_\mathrm{orbit}=\frac{\boldsymbol L^2}{2m_\mathrm{red} r^2},
\end{equation}
where the reduced mass is equal to
$$\textstyle
m_\mathrm{red}=\frac{1}{2}\,m_\mathrm{const}\approx \frac{1}{4}\,E_\mathrm{bag}.$$ At
$L=1$ and $L\sim m_\mathrm{red}\,r$, we find
$$
    E_\mathrm{orbit}\sim
    m_\mathrm{red}\approx
    \frac{1}{4}\,E_\mathrm{bag}\approx 155\mbox{ MeV.}
$$
Therefore, the mass of low-lying excitation of bag is equal
to\footnote{One could get the similar estimate, if one starts from
the calculation of mass for the $P$-wave state of D-meson, so that
in the same approximation we obtain
$m_\mathrm{D}(1P)=m_c+\frac{1}{2}\,E_\mathrm{bag}^\prime+E_\mathrm{orbit}^\mathrm{D}$,
where $E_\mathrm{orbit}^\mathrm{D}\approx
m_\mathrm{red}^\mathrm{D}=m_c
m_\mathrm{const}/(m_c+m_\mathrm{const})\approx 260$ MeV, while the
experimental data are not complete in order to calculate the
position of spin-average level, but we could hold the trend by
setting $m_\mathrm{D}^\mathrm{exp}(1P)\approx(5
m_\mathrm{D^*_2}+3m_\mathrm{D_1})/8\approx 2445$ MeV.}
\begin{equation}\label{bag'}
    E_\mathrm{bag}^\prime\approx 1105\mbox{ MeV.}
\end{equation}
Then, we easily determine the increment of repeating the low-lying
states in the system of bag-charmonium
\begin{equation}\label{increm}
    \delta
    M_\mathrm{bag}=E_\mathrm{bag}^\prime-E_\mathrm{bag}\approx
    490\mbox{ MeV,}
\end{equation}
so that we get the prediction permitting the direct verification
by the comparison with known exotic states of charmonium family,
\begin{equation}\label{predict'}
\begin{array}{ccl}
    M_\mathrm{bag}^\mathrm{c\bar c}[1^3S_1]^\prime\approx
    4260\mbox{ MeV} &\mapsto& Y(4260),\\[2mm]
    M_\mathrm{bag}^\mathrm{c\bar c}[2^3S_1]^\prime\approx
    4366\mbox{ MeV} &\mapsto& Y(4360).
\end{array}
\end{equation}

The next excitation of bag is the straightforward analog by the
analysis of $\pi$ and $\rho$ system, namely, by considering the
radial excitation $2S$: $\pi(1300)$ and $\rho(1450)$. Then, in
accordance with the experimental data\footnote{The mass of
$\pi(1300)$ is measured with uncertainty of 100 MeV.}
\begin{equation}\label{double-prime}
    E_\mathrm{bag}^{\prime\prime}\approx 1420\pm 25\mbox{ MeV.}
\end{equation}
Therefore, the masses of vector states in such the bag are equal
to
\begin{equation}\label{bag"}
\begin{array}{ccl}
    M_\mathrm{bag}^\mathrm{c\bar c}[1^3S_1]^{\prime\prime}\approx
    4575\pm25\mbox{ MeV} &\mapsto& ?,\\[2mm]
    M_\mathrm{bag}^\mathrm{c\bar c}[2^3S_1]^{\prime\prime}\approx
    4680\pm25\mbox{ MeV} &\mapsto& Y(4660).
\end{array}
\end{equation}

Thus, the model of bag-charmonium allows us to totally describe
the wide spectrum of experimental data on the spectroscopy of
extraordinary states in the charmonium family by following the
qualitatively clear assumptions and fixing the parameters from the
measured splitting between the vector and pseudovector states.

\subsection{The spectrum of charmonium family}

\begin{figure*}[th]
  \setlength{\unitlength}{1.2mm}
  \begin{center}
  \begin{picture}(139,120)
  \put(0,0){
  \includegraphics[width=130\unitlength]{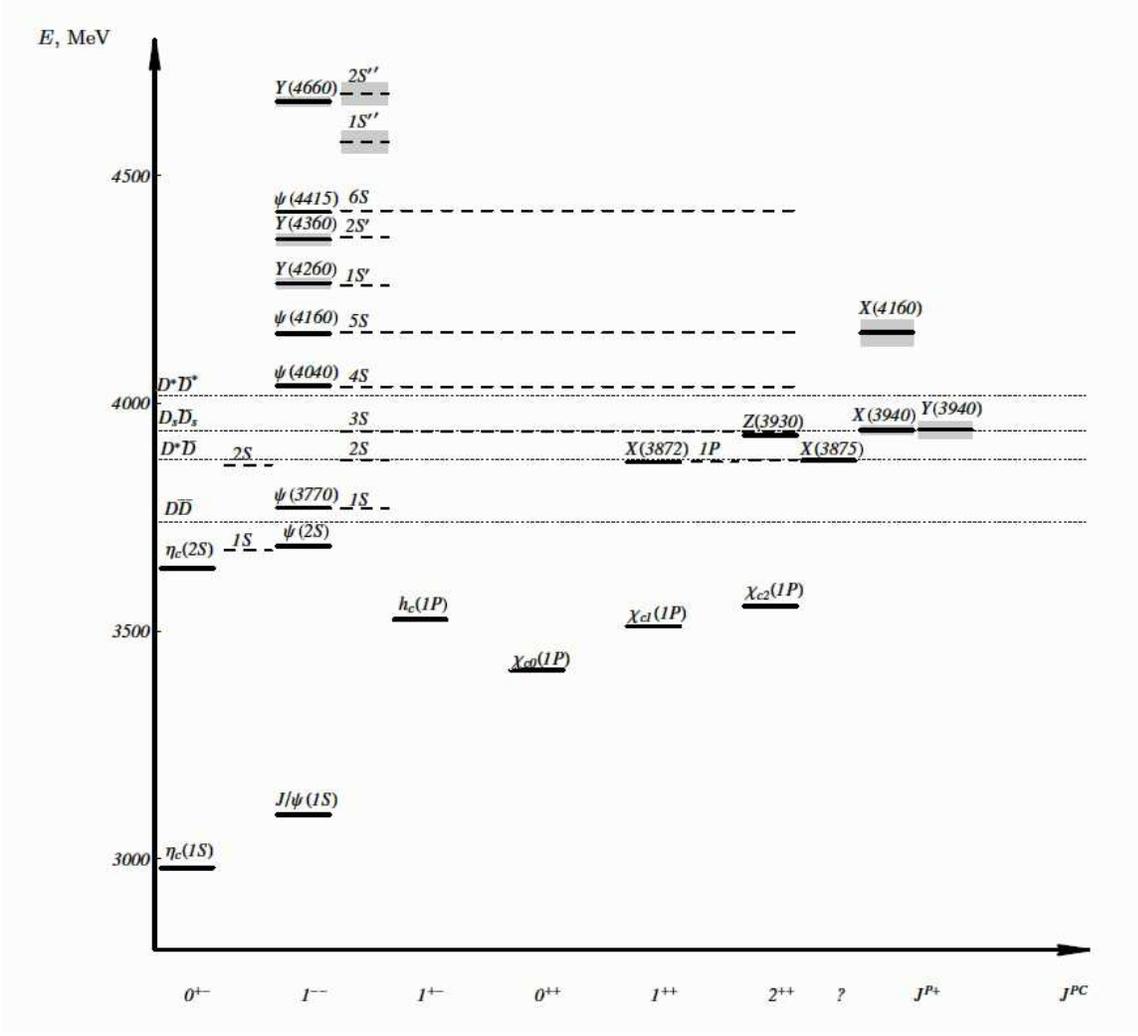}}
  \end{picture}
  \end{center}
  \caption{The spectrum of charmonium family arranged by the
  quantum numbers $J^{PC}$: solid lines mark the experimental data \cite{PDG};
  dashed lines show the predictions in the model of bag-charmonium
  (see details in the text); dotted lines are positioned at the
  thresholds for the production of mesons with the open charm. The
  shaded bands give uncertainties in the case of both experimental
  data and theoretical predictions.}\label{spectr}
\end{figure*}

The final spectrum of states for the bag-charmonium is shown in
Fig. \ref{spectr} in comparison with the experimentally measured
positions of levels in the charmonium family including the extra
states arranged by the total momentum, spatial and charged
parities, $J^{PC}$, if the quantum numbers are empirically
established or in the case of bag-charmonium with a sizable
splittings depending on the spin, i.e. at $n=1,2$. The higher
excitations at $n>2$ are marked by the dashed line drawn in the
limits from $1^{--}$ to $2^{++}$ in order to clearly point to the
fact that such the prediction concerns for the permitted higher
values of orbital momentum, not only the vector states. In that
case one has to take into attention that the $S$-wave label $nS$
refers to the pseudoscalar and vector mesons, only.

The levels of bag-charmonium are marked by the radial and orbital
quantum numbers: $n_rl$. The prime in the notation means that the
bag has got its first $P$-wave excitation, and the double prime
points to the first $S$-wave excitation of bag.

In favor of state identification accepted, there are the following
facts:
\begin{itemize}
    \item the anomaly in the distribution over the invariant mass
    of two pions in the decay $\psi(3770)\to J/\psi \pi\pi$,
    \item the preference (or the evident selectivity) in two-pion transitions
    of $Y(4260)\to J/\psi \pi\pi$ and $Y(4360)\to
    \psi^\prime\pi\pi$ for the $1S$- and $2S$-states of
    charmonium, respectively,
    \item the analogous selectivity in the two-pion transition of
    $Y(4660)\to\psi^\prime\pi\pi$ for the certain $2S$-level of charmonium.
\end{itemize}

\section{Two-pion transitions\label{III}}

The interaction of compact color-singlets composed of heavy quark
and antiquark, with ``soft degrees of freedom'', for instance,
with the low energy pions, is constructed in the framework of
multipole expansion of QCD \cite{multipoleQCD}, so that in the
leading order over the heavy quark velocity the transition of
bag-quarkonium to the state of quarkonium with the emission of two
pions posed in the isospin-singlet and $S$-wave state, takes place
due to the second order over the chromoelectric dipole moment with
the amplitude\footnote{See the methodic presentation in review
\cite{Voloshin-rev}.}
$\mathcal{A}[\mathrm{bag\mbox{-}(\hskip-1ptQ\bar Q)}_1\to
\mathrm{(\hskip-1ptQ\bar Q)}_2\pi\pi]\equiv \mathcal{A}$ equal to
\begin{equation}\label{amp1}
     \mathcal{A}=
    \langle\pi\pi\mathrm{(\hskip-1ptQ\bar Q)_2}|g_s^2
    \mathcal{E}^a_\gamma\mathcal{E}^b_\beta\,
    r_\gamma\, T^a\mathcal{G}\,T^b\,r_\beta|\mathrm{bag\mbox{-}
    (\hskip-1ptQ\bar Q)}_1\rangle,
\end{equation}
where $T^a$ is the ``difference of generators for the color
charges'' of quark and antiquark:
$T^a=\frac{1}{2}(t^a_\mathrm{Q}-t^a_\mathrm{\bar Q})$ in terms of
Gell-Mann matrices $\lambda^a=2t^a$, $\mathcal{G}$ denotes the
Green function of color-octet state of quark and antiquark, while
subscripts of quark-antiquark state mark the set of its quantum
numbers. In matrix element (\ref{amp1}) one could isolate two
following contributions: the first corresponds to the
factorization of soft degrees of freedom
$\mathcal{A}[\mathrm{bag\mbox{-}(\hskip-1ptQ\bar Q)}_1\to
    \mathrm{(\hskip-1ptQ\bar Q)}_2\pi\pi]\Big|_\mathrm{SF}=
    \mathcal{A}_\mathrm{SF}$, that takes place in
the leading order over the velocity of heavy quarks,
\begin{equation}\label{amp-soft}
\begin{array}{rcl}
    \mathcal{A}
_\mathrm{SF}&=&
    \langle\pi\pi|\,g_s^2
    \mathcal{E}^a_\gamma\mathcal{E}^b_\beta\,|\mathrm{bag}\rangle
    \,\raisebox{1pt}{$\scriptstyle\times$}
    \\[2mm]
    &&\langle
    \mathrm{(\hskip-1ptQ\bar Q)_2}|\,
    r_\gamma\, T^a\mathcal{G}T^b\,r_\beta\,|\mathrm{
    (\hskip-1ptQ\bar Q)}_1\rangle,
\end{array}
\end{equation}
so that one could introduce the chromoelectric polarizability  in
the transition between the color-singlet states of heavy quark and
antiquark
\begin{equation}\label{chromo-a}
    \alpha_{\gamma\beta}^{(12)}=\frac{1}{4}\,\langle
    \mathrm{(\hskip-1ptQ\bar Q)_2}|\,
    r_\gamma\, T^a\mathcal{G}T^a\,r_\beta\,|\mathrm{
    (\hskip-1ptQ\bar Q)}_1\rangle,
\end{equation}
which gives
\begin{equation}\label{amp-soft2}
    \mathcal{A}
_\mathrm{SF}=
    \frac{1}{2}\,\alpha_{\gamma\beta}^{(12)}\,
    \langle\pi\pi|\,g_s^2
    \mathcal{E}^a_\gamma\mathcal{E}^a_\beta\,|\mathrm{bag}\rangle.
\end{equation}
The second contribution appears due to account for the bag
annihilation $\mathcal{A}[\mathrm{bag\mbox{-}(\hskip-1ptQ\bar
Q)}_1\to \mathrm{(\hskip-1ptQ\bar Q)}_2\pi\pi]\Big|_\mathrm{BA}=
\mathcal{A}_\mathrm{BA},$ which becomes possible in higher order
over the heavy quark velovity\footnote{In addition, the
propagation of color-octet state of quarkonium with subtractive
forces suggests that the wave functions of quarks are essentially
overlapped with the bag, that can result in the annihilation of
bag.},
\begin{equation}\label{amp-annih}
\begin{array}{rcl}
    \mathcal{A}
_\mathrm{BA}&=&
    \langle\pi\pi|\,g_s^2
    \mathcal{E}^a_\gamma\mathcal{E}^b_\beta\,|0\rangle
    \raisebox{1pt}{$\scriptstyle\times$}
    \\[2mm]
    &&\langle
    \mathrm{(\hskip-1ptQ\bar Q)_2}|\,
    r_\gamma\, T^a\mathcal{G}T^b\,r_\beta\,|\mathrm{bag\mbox{-}
    (\hskip-1ptQ\bar Q)}_1\rangle,
\end{array}
\end{equation}
that is also reduced to the introduction of analog to the
chromoelectric polarizability
\begin{equation}\label{chromo-b}
    \tilde\alpha_{\gamma\beta}^{(12)}=\frac{1}{4}\,\langle
    \mathrm{(\hskip-1ptQ\bar Q)_2}|\,
    r_\gamma\, T^a\mathcal{G}T^a\,r_\beta\,|\mathrm{bag\mbox{-}
    (\hskip-1ptQ\bar Q)}_1\rangle,
\end{equation}
which can be nonzero beginning from the second order over the
heavy quark velocity, since in the first order over the velocity
the operators of interaction (dipole moments) take nonzero color
charge, $\tilde \alpha\sim \mathcal{O}(v_\mathrm{Q}^2)$. The
restriction to the $S$-wave vector states of quark and antiquark
with polarization vectors $\boldsymbol \epsilon_{1,2}$ immediately
gives
$$
    \alpha_{\gamma\beta}=\alpha\,\delta_{\gamma\beta}\;(\boldsymbol
    \epsilon_1\cdot\boldsymbol \epsilon_2^*),
$$
hence,
\begin{equation}\label{amp2}
\begin{array}{rcl}
    \mathcal{A}&=&
    \frac{1}{2}\,\alpha^{(12)}\,
    \langle\pi\pi|\,g_s^2
    \mathcal{E}^a_\beta\mathcal{E}^a_\beta\,|\mathrm{bag}\rangle
    \\[2mm]&+&
    \frac{1}{2}\,\tilde\alpha^{(12)}\,
    \langle\pi\pi|\,g_s^2
    \mathcal{E}^a_\beta\mathcal{E}^a_\beta\,|0\rangle.
\end{array}
\end{equation}
The standard technique \cite{Novikov-Shifman}, presented, for
instance, in review \cite{Voloshin-rev}, allows us to relate the
operator quadratic in the chromoelectric field with the anomaly in
the trace of energy-momentum tensor in QCD, so that in the chiral
limit \cite{chiral} we get
\begin{equation}\label{soft-3}
    \frac{1}{2}\,\tilde\alpha^{(12)}\,
    \langle\pi\pi|\,g_s^2
    \mathcal{E}^a_\beta\mathcal{E}^a_\beta\,|0\rangle=-\frac{4\pi^2}{\beta_0}\,
    \tilde\alpha^{(12)}\,q^2,
\end{equation}
where $q=p_1+p_2$ is the sum of 4-momenta of pions in the final
state, and $q^2$ denotes their invariant mass squared. Analogous
speculations in combination with an expansion of ``soft''
nonstationary state of bag over the pion states with positive
spatial and charge parities and zero charges with respect to gauge
interactions lead to the conclusion that in the limit of soft
chiral pions the dominant contribution in the matrix element is
given by the two-pion projection of bag state and, hence,
four-pion vertex, i.e. the constant integrated over the parameters
of expansion for the bag, that is approximated by
\begin{equation}\label{soft-4}
    \frac{1}{2}\,\alpha^{(12)}\,
    \langle\pi\pi|\,g_s^2
    \mathcal{E}^a_\beta\mathcal{E}^a_\beta\,|\mathrm{bag}\rangle=-\frac{4\pi^2}{\beta_0}\,
    \alpha^{(12)}\,\mu_0^2\,\mathrm{e}^{\mathrm{i}\delta},
\end{equation}
where $\delta$ defines the complex phase of matrix element
(\ref{soft-4}) with respect to (\ref{soft-3}), while the
dimensional parameter is determined by the bag mass $\mu_0\sim
E_\mathrm{bag}$, i.e. it has the magnitude of the order of several
$\Lambda_\mathrm{QCD}$. Thus, the two-pion transitions between the
vector $S$-wave states of quarkonium in the presence of bag in the
initial state are described by the functional dependence
\begin{equation}\label{amp3}
    \mathcal{A}[\mathrm{bag\mbox{-}(\hskip-1ptQ\bar Q)}_1\to
    \mathrm{(\hskip-1ptQ\bar Q)}_2\pi\pi]\sim q^2+\mu^2\,
    \mathrm{e}^{\mathrm{i}\delta},
\end{equation}
wherein the bag causes the modification by introducing the term
with $\mu\neq 0$.

It is well known that the parametrization\footnote{The standard
parametrization includes the modification allowing us to take into
account the fine effect caused by the rescattering of pions in the
final state \cite{Moxhay}, that produces corrections, which
accuracy is graded by the empirical uncertainty of real data on
the spectrum of distribution over the invariant mass of pions
\cite{cleo}.} of (\ref{amp3}) is quite accurately describes the
distribution of pions over their invariant mass in the analogous
transition $\Upsilon(3S)\to \Upsilon\pi\pi$. At first, this fact
signals that $\Upsilon(3S)$ can be treated as the basic vector
state of bag-bottomonium. Second, it shows that fitting the
parameters in the second term of (\ref{amp3}) certainly allows us
to describe the anomalous distribution in the transition of
$\psi(3770)\to J/\psi\pi\pi$, too. In this way, the value of $\mu$
is close to 0.7 GeV in the case of $\Upsilon(3S)$, that is
consistent with the prediction in the model of bag-quarkonium. On
the other hand, the appearance of such the scale in the
transitions between the ordinary states of heavy quarkonium with
the emission of two pions seems to be rather problematic. Though,
 at present, there are various versions for the
explanation of anomaly in question, which are not reduced to our
suggestion, of course (see, for example, \cite{Simonov}).

Thus, the anomalous distribution over the invariant mass of two
pions in the transition of bag-quarkonium into the heavy
quarkonium is consistent with the accepted identification of basic
vector states of bag-charmonium and bag-bottomonium.

\section{Leptonic constants\label{IV}}

The leptonic constants of heavy quarkonium can be calculated in
the potential model by applying the effective lagrangian of
nonrelativistic QCD (NRQCD), that was studied in \cite{Kis-Pakh}
in the case of heavy quark and antiquark of the same flavor and in
\cite{Bc-lept} for the quarkonium composed of heavy quark and
antiquark of different flavors. So, the vector currents of quarks
$Q$ in nonrelativistic QCD are related with the currents in the
full theory by the formula
$$
J_\nu^{QCD}= \bar Q \gamma_\nu Q, \;\;\; {\cal J}_\nu^{NRQCD} =
\chi^\dagger \sigma_\nu^\perp \phi,
$$
where $Q$ is the field of relativistic quark, while $\chi$ and
$\phi$ demote the nonrelativistic Pauli spinors for the quark and
antiquark, $\sigma_\nu^\perp= \sigma_\nu -v_\nu (\sigma \cdot v)$,
and $v$ is the 4-velocity of quarkonium, so that
\begin{equation}
J_\nu^{QCD} = {\cal K}(\mu_{\rm hard}; \mu_{\rm fact})\cdot {\cal
J}_\nu^{NRQCD}(\mu_{\rm fact}). \label{match}
\end{equation}
Here $\mu_{\rm hard}$ fixes the point of matching NRQCD to QCD,
$\mu_{\rm fact}$ denotes the scale of perturbative calculations in
 NRQCD.

In the case of quarks of the same flavor, the Wilson coefficient
${\cal K}$ is known to the 2-loop accuracy
\cite{HT,bensmir,melch,mel}
\begin{equation}
\begin{array}{rcl}
{\cal K}(\mu_{\rm hard}; \mu_{\rm fact})
&\hskip-5pt=&\hskip-3pt\displaystyle 1 -\frac{8}{3}
\frac{\alpha_s^{\overline{\Rsub MS}}(\mu_{\rm
hard})}{\pi}\\[3mm]
&\hskip-5pt+&\hskip-5pt\displaystyle
\left(\frac{\alpha_s^{\overline{\Rsub MS}}(\mu_{\rm
hard})}{\pi}\right)^2 c_2(\mu_{\rm hard}; \mu_{\rm fact}),
\end{array}\label{kfact}
\end{equation}
$c_2$ is explicitly given in \cite{bensmir,melch}. The anomalous
dimension of ${\cal K}(\mu_{\rm hard}; \mu_{\rm fact})$
\begin{equation}
\frac{d \ln{\cal K}(\mu_{\rm hard}; \mu)}{d \ln \mu} =
\sum_{k=1}^{\infty} \gamma_{[k]}
\left(\frac{\alpha_s^{\overline{\Rsub MS}}(\mu)}{4\pi}\right)^k,
\label{anom}
\end{equation}
in two loops is given by the expression\footnote{In terms of
ordinary notations for the representations of $SU(N_c)$ group:
$C_F=\frac{N_c^2-1}{2 N_c}$, $C_A= N_c$, $T_F = \frac{1}{2}$.}
\begin{eqnarray}
\gamma_{[1]} & = & 0,\\
\gamma_{[2]} & = & -16 \pi^2 C_F \left(\frac{1}{3} C_F +
\frac{1}{2} C_A\right). \label{g2}
\end{eqnarray}
Initial data for the evolution of ${\cal K}(\mu_{\rm hard};
\mu_{\rm fact})$ versus the scale are determined by the matching
conditions fixed at $\mu = \mu_{\rm hard}$ \cite{bensmir,melch}.

The leptonic constant of vector state with the polarization
$\lambda$ and polarization vector $\epsilon_\nu^\lambda$ in full
QCD
\begin{equation}
\langle 0| J_\nu^{QCD} |\mathrm{Q\bar Q} ,\lambda \rangle =
\epsilon_\nu^\lambda f_{\mathrm{Q\bar Q} } M_{\mathrm{Q\bar Q} },
\end{equation}
is related with the matrix element of current ${\cal
J}_\nu^{NRQCD}$
\begin{equation}
\langle 0| {\cal J}_\nu^{NRQCD}(\mu) |\mathrm{Q\bar Q} ,\lambda
\rangle = {\cal A}(\mu)\; \epsilon_\nu^\lambda f_{\mathrm{Q\bar Q}
}^{NRQCD} M_{\mathrm{Q\bar Q} }, \label{a}
\end{equation}
where the potential model of quarkonium with the wave function
$\Psi_\mathrm{Q\bar Q}(\boldsymbol r)$ gives
\begin{equation}
f_{\mathrm{Q\bar Q}}^{NRQCD} = \sqrt{\frac{12}{M}}\;
|\Psi_{\mathrm{Q\bar Q}}(0)|, \label{wave}
\end{equation}
while the renormalization group factor ${\cal A(\mu)}$ equals unit
at the scale of $\mu=\mu_0$, whereat the wave function is
determined. Since
\begin{equation}\label{con1}
f_{\bar QQ} = f_{\bar QQ}^{NRQCD} {\cal A}(\mu_{\rm fact})\cdot
{\cal K}(\mu_{\rm hard}; \mu_{\rm fact}), \label{cc}
\end{equation}
the anomalous dimension of ${\cal A}(\mu_{\rm fact})$ compensates
the anomalous dimension of ${\cal K}(\mu_{\rm hard}; \mu_{\rm
fact})$, hence,
\begin{equation}
\frac{d \ln{\cal A}(\mu)}{d \ln \mu} = - \gamma_{[2]}
\left(\frac{\alpha_s^{\overline{\Rsub MS}}(\mu)}{4\pi}\right)^2,
\label{anoma}
\end{equation}
therefore,
\begin{equation}
{\cal A}(\mu) = {\cal A}(\mu_0)\; \left[ \frac{\beta_0+\beta_1
{\displaystyle\frac{\alpha_s^{\overline{\Rsub
MS}}(\mu)}{4\pi}}}{\beta_0+\beta_1 {\displaystyle
\frac{\alpha_s^{\overline{\Rsub MS}}(\mu_0)}{4\pi}}}
\right]^{\displaystyle\frac{\gamma_{[2]}}{2\beta_1}}.
\end{equation}

The calculation of leptonic constants in wide limits for the
scales of matching and factorization allows us to find the region,
wherein the result is stable, i.e. it slightly depends on small
variations of scales. Then, $\mu_0$ is set to the point of
stability. Numerical estimates are consistent with the
experimental data for the basic vector states of charmonium and
bottomonium \cite{Kis-Pakh}, while the leptonic constants of
excited states in such the approach are simply determined from the
ratio of wave functions at the origin for the given state to the
wave function of basic state due to appropriate rescaling of
leptonic constant for the basic state in accordance with
(\ref{con1}), since the renormalization group factors are
universal , i.e. they do not depend on the excitation number.
Then, the problem is reduced to the safe calculation of wave
functions of quarkonium at the origin. Such the calculation
involves rather a large uncertainty because of modelling the
potential and prescribing the pole masses of quarks depending on
the model. Thus, one arrives to a sizable methodic uncertainty of
theoretical predictions. Nevertheless, one can reliably state that
the low value of leptonic constant of $\psi(3770)$ is caused by
the fact that this state cannot be assigned to the $S$-wave state
of charmonium. Therefore, it is the exotic state whether it is the
result of mixing the $S$-wave level with the $D$-wave one, as
usually accepted, in spite of the fact that the splitting between
such the nearest levels is about 100 MeV, and hence, the matrix
element of mixing should get the same order of magnitude, that is
quite a large value.

However, such the framework is not applicable to the states of
bag-quarkonium, since in this case the joint annihilation of
quark-antiquark pair and bag should take place. Then, the
knowledge of wave function for the quark-antiquark system is not
sufficient. In addition, the wave function could be essentially
modified during the annihilation with the bag. So, let us explore
the method of quasilocal sum rules of QCD \cite{quasilocal-SR}.
The basic idea of such the approach is reduced to the following:
\begin{itemize}
    \item Consider the transversal part of correlator for the
    vector currents of heavy quarks, namely, the real part of
    correlator at the invariant mass set to zero as well as the
    derivatives of $k$-th order, i.e the moments of
    real part at zero.
    \item In the case of heavy quarks, the bound states are
    positioned in the narrow gap which width is suppressed in
    comparison with  the heavy quark mass $m_\mathrm{Q}$, so that
    for the low values of moment numbers one can transform the
    summation over the resonances to the integration over the
    density of states
    $\mathrm{d}\tilde n/\mathrm{d}M_{\tilde n}$, where $\tilde n$
    denotes the number of vector state in the direction of mass
    increase\footnote{We insert the tilde symbol in order to
    distinguish the number from the principal quantum number $n$ used
    above.}, with accuracy up to $\mathcal{O}(1/m_\mathrm{Q})$.
    \item In the leading order over the inverse mass $1/m_\mathrm{Q}$
    at low values of moment number,
    calculating the correlator can be restricted by the
    contribution of nonrelativistic quarks with account for their
    coulomb-like interaction.
\end{itemize}
Then, at the points of resonances the local equality of
theoretical correlator to the empirical one approximated due to
the introduction of density of states leads to the relation for
the leptonic constant of $\tilde n$-th state $f_{\tilde n}$
\begin{equation}\label{quasilocal}
    \frac{f_{\tilde n}^2}{M_{\tilde n}}=\mathcal{C}\,
    \frac{\mathrm{d}M_{\tilde n}}{\mathrm{d}\tilde n},
\end{equation}
where the constant is defined by the formula
\begin{equation}\label{C}
    \mathcal{C}=\frac{\alpha_s^\mathrm{c}}{\pi}\,\mathcal{K}^2\left(
    \frac{2m_\mathrm{Q}}{M_{\tilde n}}\right)^2\,Z_\mathrm{sys},
\end{equation}
with an effective constant of coulomb exchange\footnote{The
effective constant of coulomb exchange as well as the quark mass
in this approximation do not necessary coincide with their analogs
in the framework of potential approach. Though, the mentioned
differences become less with the increase of quark mass, so that
for the bottomonium these constants coincide with the accuracy
below 10\%, indeed, if one does not take into account for the
difference in the values of $b$-quark pole mass, while the account
for the renormalization group evolution of coupling constant
versus the characteristic momentum inside the quarkonium, hence,
versus the quark mass, leads to that the effective constants
actually coincide with each other with the accuracy better than
2\%, that is essentially underestimate the accuracy of assumptions
made in the way of deriving these estimates. In addition, the
strict analysis of charmonium and bottomonium made in the
framework of assumption on the dominant coulomb interaction with
account for the perturbative and nonperturbative corrections in
\cite{TY1,TY2,Y3,Y+} led to the values of pole masses of charmed
and bottom quarks, which are consistent with the values obtained
in the present paper within the accuracy of estimates. The
effective constants of coulomb interaction take similar values,
too.} $\alpha_s^\mathrm{c}$, the factor of loop corrections
$\mathcal{K}$ taken to the 1-loop accuracy from (\ref{kfact}) and
a systematic correction generally depending on the moment number
$$
    Z_\mathrm{sys}=\frac{Z_\mathrm{NR}}{Z_\mathrm{int}},
$$
introduced in the both nonrelativistic approximation of quark
correlator and density of states for the integral representation
for the sum over the resonances. In the limit of heavy quarks,
i.e. by neglecting logarithmic and power corrections of the form
$\ln \mu/m_\mathrm{Q}$ and $1/m_\mathrm{Q}$, the effective
constant $\mathcal{C}$ is independent of quark flavor. It is
spectacular, that empirically for the finite masses of real $c$-
and $b$-quarks and basic vector states, we have got\footnote{The
exact value can depend on the choice of quark mass, that leads
also to a variation of systematic factor caused by the
nonrelativistic approximation in the calculation of coulomb
contribution near the threshold.}
$$
    \mathcal{C}\approx\frac{1}{5\pi},
$$
so that the scaling relation of (\ref{quasilocal}) takes place.

If the heavy quark and antiquark states are certainly the levels
of nonrelativistic heavy quarkonium, then the density of states
is, in practice, universal, since the potential is close to the
logarithmic dependence, so that independently of the heavy quark
flavor we get
\begin{equation}\label{density}
    \frac{\mathrm{d}M_{\tilde n}}{\mathrm{d}\tilde n}\approx \frac{2 T}{\tilde n},
\end{equation}
where $T$ is the average kinetic energy, i.e. the main parameter
of logarithmic potential. Therefore, for the heavy quarkonium the
scaling law takes place
\begin{equation}\label{scaling}
    \frac{f_{\tilde n}^2}{M_{\tilde n}}\,\tilde n=\mbox{const.},
\end{equation}
valid for $J/\psi$ and $\Upsilon$. By the way, since the width of
vector quarkonium decay $\Gamma[1^{--}_\mathrm{Q\bar Q}\to
e^+e^-]$ is related with the electric charge of quark
$e_\mathrm{Q}$ and its leptonic constant by the expression
$$
    \Gamma_{\tilde n}=\frac{4\pi}{9}\,\alpha_\mathrm{em}^2(M_{\tilde n})\,
    e^2_\mathrm{Q}\,\frac{f_{\tilde n}^2}{M_{\tilde n}},
$$
for the standard states of heavy quarkonium with the regular
spectrum under condition (\ref{density}) one could expect the
universal law
\begin{equation}\label{gama-law}
    \frac{1}{e^2_\mathrm{Q}}\,\Gamma_{\tilde n}\,\tilde n=\mbox{const.}
\end{equation}

Relation (\ref{quasilocal}) is also applicable, if there are
exotic, extraordinary states with the hidden charm, too. Its
physical meaning is simple: the contributions of resonances into
the correlator are actually fixed by the dislocation in accordance
with the \textit{observed} density of states. Therefore, it is
enough to carry out the analysis of data on the spectrum of vector
states with the hidden charm and, then, to test the direct
consequence of (\ref{quasilocal}) in the form of
\begin{equation}\label{quasi-lept}
    \frac{1}{e^2_\mathrm{Q}}\,\Gamma_{\tilde n}\,
    \frac{\mathrm{d}\tilde n}{\mathrm{d}M_{\tilde n}}=\mbox{const.}
\end{equation}

\begin{figure}[th]
  \setlength{\unitlength}{1.mm}
  \begin{center}
  \begin{picture}(85,60)
  \put(3,3){
  \includegraphics[width=80\unitlength]{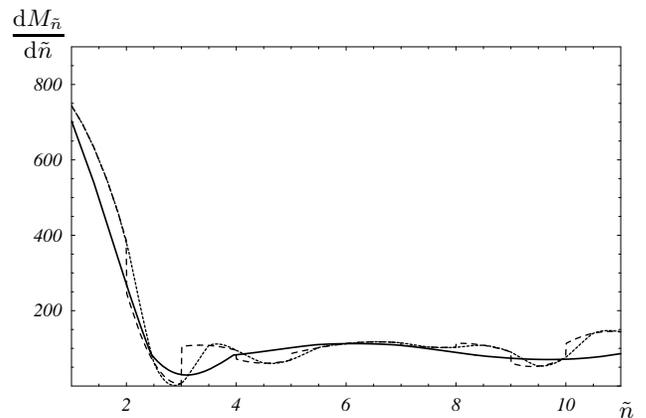}}
  \put(83,3){$\tilde n$}
  \put(2,53){$\displaystyle \frac{\mathrm{d}M_{\tilde n}}{\mathrm{d}\tilde n}$}
  \end{picture}
  \end{center}
  \caption{The result of cubic interpolation of vector state
  spectrum of charmonium family in accordance with predictions of
  model with the consequent calculation of
  $\mathrm{d}M_{\tilde n}/\mathrm{d}\tilde n$ (dashed line) in
  comparison with the smoothed cubic interpolation of the state
  density involving the increment steps equal to 1.5 and
  0.8 in $\tilde n$ (solid and dotted lines, correspondingly).
  }\label{cubic}
\end{figure}

The accuracy of analyzing the spectrum of vector states is
essentially restricted by the fact that the levels are discrete.
The result of cubic interpolation for the mass spectrum with the
further calculation of $\mathrm{d}M_{\tilde n}/\mathrm{d}\tilde n$
versus $\tilde n$ being the number of vector state as given in
Fig. \ref{spectr} by the mass increase, is shown in Fig.
\ref{cubic}.

Note that the interpolation of spectrum itself versus the
resonance number leads to discontinuities of the mass derivative
with respect to the number exactly at the points, which refer to
the values of state density in interest. Therefore, we have made
the cubic interpolation of the state density itself once more with
different increments in the number in order to smoothen the
discontinuities\footnote{The artefact of function discontinuity
points to the fact that the derivative of spectrum with respect to
the number of resonance permits a displacement of point on the
number axis in the limits $\delta n=0.5$, so that we have also
applied the method of smoothing by averaging the values of
spectrum density taken left and right in vicinity of resonance,
that agrees with the uncertainties of estimation itself. Then, the
smoothing is slowly modify the values of spectrum density in the
points, where it fluctuates weakly, i.e. the spectrum is rather
regular, and in addition, it allows us to draw the limits of
approach uncertainties.} and show uncertainties of estimating the
quantity of $\mathrm{d}M_{\tilde n}/\mathrm{d}\tilde n$. In figure
we see that the most reliable results are given at $\tilde n=6,7$
(states $\psi(4040)$ and $\psi(4156)$), whereat the spectrum is
the most regular in the sense of its smoothness by incrementing
between the nearest states. The uncertainty grows at $\tilde n=3$
(the state $\psi(3770)$), whereat the mass increment is extremely
inhomogeneous, as well as at the end of interpolation interval,
i.e. at the important positions of states with $\tilde n=1,2$,
mesons $J/\psi$ and $\psi(2S)$. The absolutely analogous situation
takes place at other choice of interpolation power, namely, at
linear and quadratic interpolations, for instance. This fact means
that the investigation of relation (\ref{quasi-lept}) should be
made by fixing the constant at $\tilde n=6$, say, that we will
perform further. In this way, an essential uncertainty in the
calculation of leptonic constant of basic state $J/\psi$ appears.
In order to decrease the uncertainty of interpolating towards
$J/\psi$, we will take into account that this state is certainly
the heavy quarkonim, hence, the state density in vicinity of
$\tilde n=1$ is rather close to the value predicted in the
potential model. This fact can be naturally involved by
introducing an auxiliary state with $\tilde n=0$ and mass
$M_{0}\approx M_{J/\psi}-2 T\approx 2296$ MeV according to the
linear extrapolation, that we actually make.

\begin{figure*}[th]
  \setlength{\unitlength}{1.5mm}
  \begin{center}
  \begin{picture}(85,56)
  \put(3,3){
  \includegraphics[width=80\unitlength]{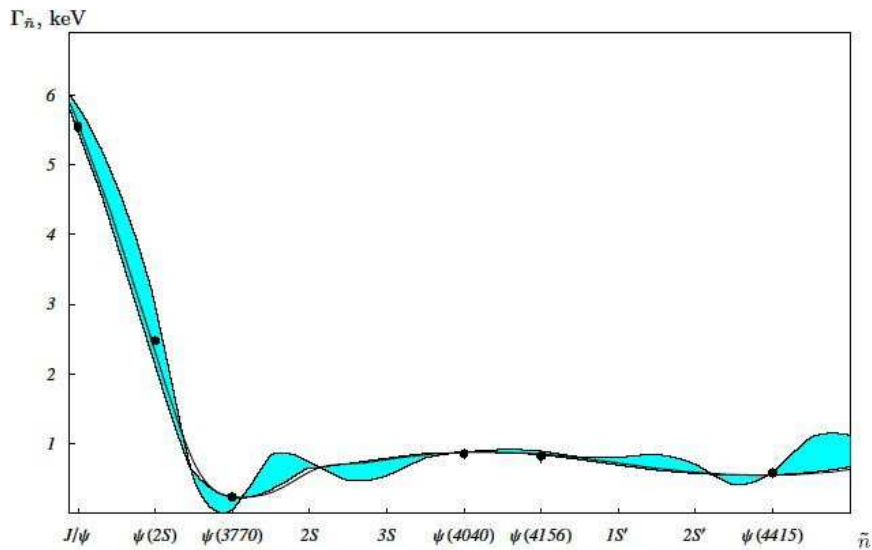}}
  \end{picture}
  \end{center}
  \vspace{-5mm}
  \caption{The leptonic constants of vector states in the family
  of charmonium $\Gamma_{\tilde n}$ in keV versus the resonance
  number $\tilde n$ (solid line). The shaded region marks the
  systematic uncertainty of predictions, which is caused by the
  interpolation of state density. The dots denote the experimental
  data with the uncertainties shown as vertical lines.
  }\label{interpolate-l}
\end{figure*}

Thus, relation (\ref{quasi-lept}) leads to estimation of leptonic
constants shown in Fig. \ref{interpolate-l}. We see that the
scaling relation quite successfully describes the behavior of
leptonic widths for the observed states. Since the relation is
universal, and it is valid irrespectively of the role played by
infrared phenomena such as the bag, there is no need to prescribe
the state of $\psi(3770)$ as the $1D$ level of charmonium with the
appropriate mixing with the $2S$-level in order to produce nonzero
value of leptonic constant (the wave function at the origin).
Moreover, the small difference between the leptonic constant of
$\psi(4040)$ and $\psi(4156)$ looks quite exotic, if one treats
them as the $S$-wave levels of charmonium with the respective
numbers $\tilde n\mapsto 3,4$, while the presence of extra states
due to the bag-charmonium leads to the both equalizing the ratio
of leptonic widths and more natural values of absolute values:
$\Gamma_6\approx\frac{1}{6}\Gamma_1$. Similar statements can be
written as concerns for the state of $\psi(4415)$.

We do not present a complete description for the procedure of
interpolating the spectrum density, since Fig. \ref{interpolate-l}
clearly shows that the method inherently involves irreducible
uncertainties, which do not permit to make the theoretical
predictions with the higher precision\footnote{For instance, we
could quite smoothly fit the spectrum of vector states with the
hidden charm by means of special set of basic functions, so that
the spectrum density will repeat the experimental dependence of
leptonic constants versus the resonance number. However, the
procedure of selecting the basic functions is arbitrary itself, so
it involves almost uncontrolled methodic uncertainty, restricted
by the constraint of acceptable smoothness of the fit, only.},
comparable with the accuracy of experiment. Nevertheless, the
exploration of quasilocal sum rules by introducing the density of
bound states with the hidden charm allows us to  establish the
tendency in the character of dependence of leptonic constants on
the number of excitation over the basic state.

Thus, the extraordinary states in the family of charmonium as
dictated by the model of bag-charmonium, do not contradict with
the analysis of know leptonic constants of vector states, and
moreover, they permit to improve the systematization of those
constants (or leptonic widths).

\section{Conclusion}

In the present paper we have formulated the potential model of
bag-charmonium, which has allowed us to successfully identify
extraordinary states in the family of charmonium in terms of
spectroscopy, to substantiate the anomaly in the distribution over
the invariant mass of two pions in hadronic transitions such as
$\psi(3770)\to J/\psi\pi\pi$ and $\Upsilon(3S)\to\Upsilon\pi\pi$
by assigning the initial states as the vector bag-quarkonia and to
establish the systematic regularity of leptonic constants for the
vector states in the charmonium family  due to the levels of
bag-quarkonium. The regularity has got the form of scaling law
derived from the quasilocal sum rules of QCD.

However, such the model predicts some new states yet not observed
empirically.

As concerns for the mechanisms of production of exotic states, one
could stress the following: in decays of heavy hadrons with the
light valence quark, the production of bag-charmonium goes almost
the same way as the production of charmonium, namely, at rather
short distances about the charmonium size, the compact state of
charmed quark and antiquark is formed, while the bag itself
already exists due to the presence of light valence quark in the
initial state, so that the bag introduces an appropriate form
factor, describing a possible transition of bag from the rest in
the initial state to a motion joint to the quark-antiquark pair
with the hidden charm. An analogous form factor, probably, should
appear at the production of bag-charmonium in hadronic collisions,
whereas the order of magnitude for the cross section is close to
the production cross section of charmonium. In leptonic
collisions, the main role is played by the leptonic constants
studied in Section \ref{IV}.

In the present paper we, in fact, have not discussed any channels
of decays for the bag-quarkonium except pointing to the
selectivity in the two-pion transitions of bag-charmonium to the
charmonium states with identical quantum numbers of
quark-antiquark pair.

In this respect, it is appropriate once more to refer to review
\cite{Godfrey-rev}, wherein there is a comprehensive bibliograthy
of original articles concerning for the models of exotic
quarkonium states, their decays and production modes as well as a
comparative description of various mechanisms of forming the
extraordinary states in the charmonium family. We point to some
pioneer papers, wherein authors introduced such the notions into
the scientific language in the field of quark dynamics of exotic
states with the hidden charm as follows:
\begin{itemize}
 \item the hadronic molecule \cite{VO,DeRGG} or deuson
 (in analog to the deutron) \cite{Tornq}, composed of charmed
 mesons with the positive and negative charm $D\bar D$,
 \item the hybrid \cite{BT-h,BCS} containing the quark-antiquark
 pair in the color-octet, which is coupled with a compact gluonic lump
 to form the colorless state,
 \item the tetraquark \cite{Maiani-Polosa} composed of diquark and
 diantiquark.
\end{itemize}
In addition, it worth to mention the recent paper on the
systematization of heavy quarkonium states -- charmonium and
bottomonium -- in the scheme of Regge trajectories
\cite{LikhoLuch}. Studying various color structures of tetraquark
has been given in \cite{GLP}.

This paper is partially supported by RosAtom, contract
\#~N.4d.47.03.08.069, and RFBR, grant \#~07-02-00417.

\end{document}